# LUNAR LASER RANGING TEST OF THE INVARIANCE OF $c$

Daniel Y. Gezari

NASA/Goddard Space Flight Center, Laboratory for ExoPlanets and Stellar Astrophysics, Code 667, Greenbelt, MD 20771

*and*

American Museum of Natural History, Astrophysics Department, New York, NY 10024

*daniel.y.gezari@nasa.gov*

Abstract

The speed of laser light pulses launched from Earth and returned by a retro-reflector on the Moon was calculated from precision round-trip time-of-flight measurements and modeled distances. The measured speed of light ($c$) in the moving observer's rest frame was found to exceed the canonical value $c = 299,792,458$ *m/s* by $200\pm10$ *m/s*, just the speed of the observatory along the line-of-sight due to the rotation of the Earth during the measurements. This result is a first-order violation of local Lorentz invariance; the speed of light seems to depend on the motion of the observer after all, as in classical wave theory, which implies that a preferred reference frame exists for the propagation of light. However, the present experiment cannot identify the physical system to which such a preferred frame might be tied.

## 1. INTRODUCTION

We have measured the two-way speed of light ($c$) using lunar laser ranging to test the invariance of $c$ to motion of the observer, a necessary condition for the local Lorentz invariance of $c$ and the fundamental assumption underlying all of the predictions of the special theory of relativity (Einstein 1905) in the matter and photon sectors. Surprisingly, a review of the experimental literature finds no previous published report of an attempt to measure $c$ directly with a moving detector to confirm that light actually propagates this way (Gezari 2009).

The question immediately arises: Why bother to search for first-order violations of Lorentz invariance in the photon sector when sensitive modern experiments have yet to reveal any conflicting second-order optical effects? Attempts are even being made to detect third-order violations predicted by contemporary field theories, which allow for spontaneous CPT and Lorentz symmetry breaking, and even for the possibility that a preferred or absolute reference frame might exist (for overviews see Wilczek 1999, Pospelov and Romalis 2004). In fact, correlated, first-order variations in the time-of-flight of electromagnetic signals measured with moving receivers are commonly observed in pursuit of much more subtle phenomena, such as in experiments searching for evidence of micro-gravity and quantum gravity effects (*e.g.*, Williams et al. 1996, 2004) or in the operation of the Global Positioning System (GPS) satellite navigation system. These timing variations are easily detected – and routinely corrected for – in modern optical ranging experiments, however, the full significance of such correlated, first-order effects has apparently not been fully appreciated.

The local Lorentz invariance of $c$ can now only be inferred from observations of moving sources, symmetry arguments, and the null results of ether drift and speed-of-light isotropy experiments. However, there are real difficulties with this view. Observations of moving sources cannot discriminate between special relativity and the old ether hypothesis, and do not favor one over the other, because the invariance of $c$ to motion of the emitting source is a common feature of both special relativity and classical wave theory of light (discussed by Gezari 2009). Of course, one could argue that experiments with moving sources and moving observers should be equivalent and indistinguishable. But in some optical phenomena, motion of the source and motion of the observer have entirely different consequences (*e.g.*, the Doppler effect in an optical medium, stellar aberration). To claim that source motion and observer motion are equivalent without experimental confirmation would be invoking the theory to validate itself, and such experiments have not yet been performed. Furthermore, the null results of the well-known ether drift experiments (*e.g.*, Michelson and Morley 1887) have recently come into question (Consoli and Costanza 2003), and there are similar concerns for the integrity of modern resonant-cavity speed-of-light isotropy experiments (*e.g.*, Braxmaier *et al.* 2002, Muller 2005, Muller *et al.* 2007). These issues, and the present experimental basis for special relativity in the photon sector, are discussed by Gezari (2009).

From the perspective of an experimentalist and observer this is all quite troubling. Rather that infer the invariance of $c$ from indirect evidence it would be more straightforward, and more convincing, to simply measure the speed of light directly with a moving detector that was controlled or actively monitored by the observer. We have made such a measurement using the method of lunar laser ranging.

2. EXPERIMENTAL APPROACH

In this experiment laser light pulses are launched from an observatory on Earth and are returned by a retro-reflector deployed on the surface of the Moon. The rotation of the Earth carries the observer toward and away from the Moon at velocities within a range of about ±300 *m/sec* along the Earth-Moon line-of-sight. It would be reasonable to expect that the round-trip light time would change due to this motion and, in fact, it does. The measured time-of-flight of individual laser pulses varies by as much as ~3 *sec.* This effect is more than four orders of magnitude larger than the precision achievable with standard pulse timing techniques used in lunar laser ranging.

In principle, the average speed of a light pulse out and back calculated by the moving observer is simply the optical path length traveled by that pulse in the observer's rest frame, divided by the time-of-flight of that pulse from the source to the receiver, measured in the observer's rest frame. The distance between the observatory and the retro-reflector at any moment during a measurement can be calculated with ~1 meter accuracy from well-determined ephemerides of the Moon's orbit, the exact location of the observatory relative to Earth-center and of the retro-reflector relative to Moon-center, and the local sidereal time, using modeling tools described in Section 3.2. The times at which light pulses are launched from and received back at the observatory are observed directly in the observatory rest frame.

The intrinsic geometry of this lunar laser ranging observation provides four fundamental advantages that simplify analysis of the data and interpretation of the experimental results:

*a)* The retro-reflector operating in conjunction with the laser effectively comprises a light source outside the Earth's atmosphere that is controlled by the moving observer.

*b)* Because the observer, the laser light source, the detector and the timing instrumentation are all at rest in the observatory rest frame, the moving observer has exact knowledge of the time that the outgoing laser pulse is launched and the time that the reflected pulse is received, both measured in the observatory rest frame. Thus, all simultaneity concerns are avoided, which might otherwise complicate the analysis of the pulse timing data and the interpretation of the results.

*c)* The measurement is sensitive only to motion of the detector in the local Earth-Moon stationary frame, even though the observer, the source, the retro-reflector, the Earth and the Moon are all moving in one way or another in the local stationary reference frame of the "fixed stars" due to rotations and orbital motions of the Earth and Moon. Optical path length variations due to the ellipticity of the Moon's orbit and lunar libration are small and slow to the level of precision required by our experiment on the timescale of our measurements. Motion of the Earth-Moon system in orbit around the Sun would average out in a two-way measurement, and only appear as a small (~3 *m/s*) second-order residual. Therefore, in this experimental configuration, the detector is the only optical element moving in the local stationary frame that could have observable consequences for the speed of light calculated from our measurements.

*d)* The observation is essentially a measurement of the speed of light in free space because more than 99.9% of the optical path lies outside the Earth's atmosphere. But the question remains whether the physical properties of the region between the Earth and the Moon correspond closely enough to those of free space. Scintillation (scattering) only begins to have a significant affect on the speed of visible light propagating in the interstellar medium over path lengths longer than about 1 *parsec* (Fox 1962), so the effect would be completely negligible in our application. Therefore, or the purposes of this experiment, the local interplanetary medium can be considered a reasonable approximation of free space.

# 3. OBSERVATIONS

Laser light pulses were launched to the Moon from the Apache Point Lunar Laser-ranging Operation (APOLLO) facility (Murphy *et al.* 2004, 2007) installed at the 3.5-meter telescope at Apache Point Observatory (APO) on 11 November 2007. The pulses were returned by the AP15RR retro-reflector deployed on the lunar surface during the Apollo 15 mission (Bender et al. 1973). At the time these shots were launched it was just sunrise at APO (latitude $b$ = 32.605° N, longitude $\ell$ = 254.18°), which is $7.07^h$ west of the Greenwich meridian; the azimuth and altitude angles of the Moon were $\theta$ = 122°, $\phi$ = 41°, and the angle between the Earth's orbital velocity vector ($V_E$ 30 *km/s*) and the line-of-sight to the Moon was $\psi$ = 46°. The speed of the observatory along the line-of-sight changes continuously as $v_O = v_E \cos b \sin\theta \cos\phi$ during the measurements, where $v_E$ is the Earth rotation speed at the equator and $b$ is the latitude of the observatory.

## *3.1. Time Measurement*

The timing data analyzed in this study was provided by Murphy (2008). Individual laser shots were launched at intervals of 0.05 *sec* and the time-of-flight of each pulse from the Earth to the Moon and back was measured with 0.1 *ns* timing resolution. Spurious detector events and noise were reduced by rejecting any detections occurring more than ±10 *ns* from the nominal arrival time of each pulse predicted from modeled of the motions of the Earth and the Moon. A sharp increase in the number of detector events is seen when the retro-reflector is acquired (Figure 1). The majority of the events are clustered within ±1 *ns* of their predicted arrival time.

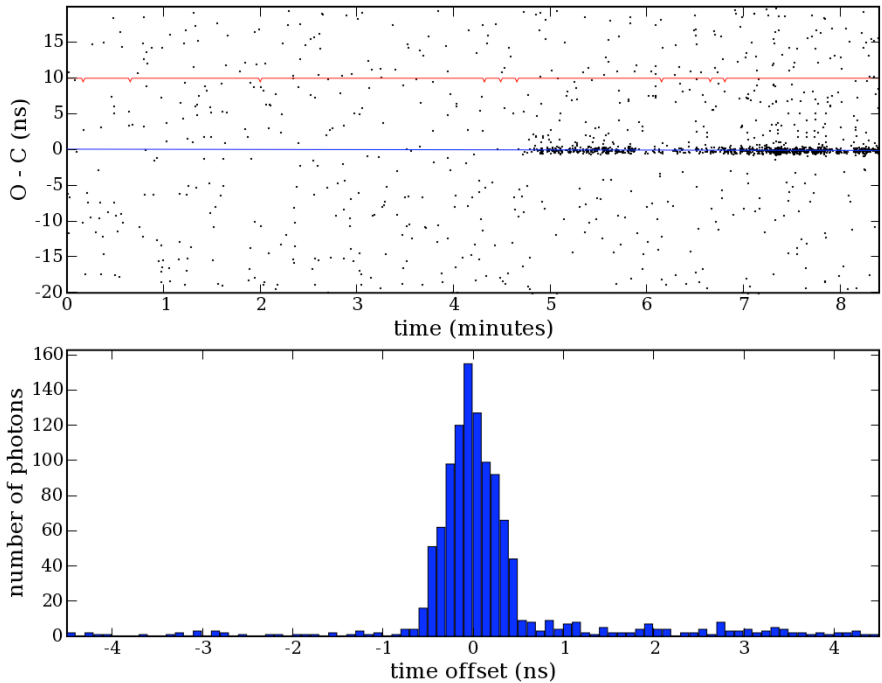

Figure 1: *Top:* Typical record detected photon events, showing the difference between predicted and actual arrival times. The onset of the detection of laser photons that were launched from the APOLLO facility and returned by the A15RR lunar retro-reflector is clearly seen at $t \sim 5$ *min*. *Bottom:* Histogram of the difference between the observed (O) and calculated (C) photon arrival times. This figure is by Murphy (2008).

## *3.2. Distance Determination*

The exact distance between the APOLLO laser light source, the A15RR retro-reflector, and the APOLLO detector was determined for each shot from the DE403 lunar ephemeris database created and maintained by NASA/Jet Propulsion Laboratory using the NAIF/SPICE Toolkit (*http://naif.jpl.nasa.gov*) and the Earth Orientation Model of the International Earth Rotation and Reference Systems Service (IERS; *http://iers.org*). These distances were calculated by modeling the Earth and Moon positions for each of the 2,636 individual detector events in the observation (Neumann 2008) at the beginning (launch) and the end (receive) of each measurement, as well as at the modeled time of reflection of the laser pulse by the retro-reflector (bounce). This approach includes corrections for vacuum light time propagation. The positions of the Earth and Moon in the ecliptic plane are calculated for each event and given $x,y$ coordinates in the solar system barycentric J2000 local stationary frame. The positions of the Earth and Moon in this system at launch, bounce and receive are illustrated in Figure 2. Note that the Earth and Moon are moving together as a binary system at ~30 *km/s* in that frame, as the Earth orbits the Sun, and relative to each other at much smaller speeds of order ~10 *m/s* due to the eccentricity of the lunar orbit. The distances between APO and A15RR are then calculated from the *x,y* positions at the times of any significant event. These distances are determined to an uncertainty of ±1*m*, which is more than adequate for our purposes since we only require moderate distance resolution (~10*m*) to obtain an unambiguous, first-order, $10\sigma$ result.

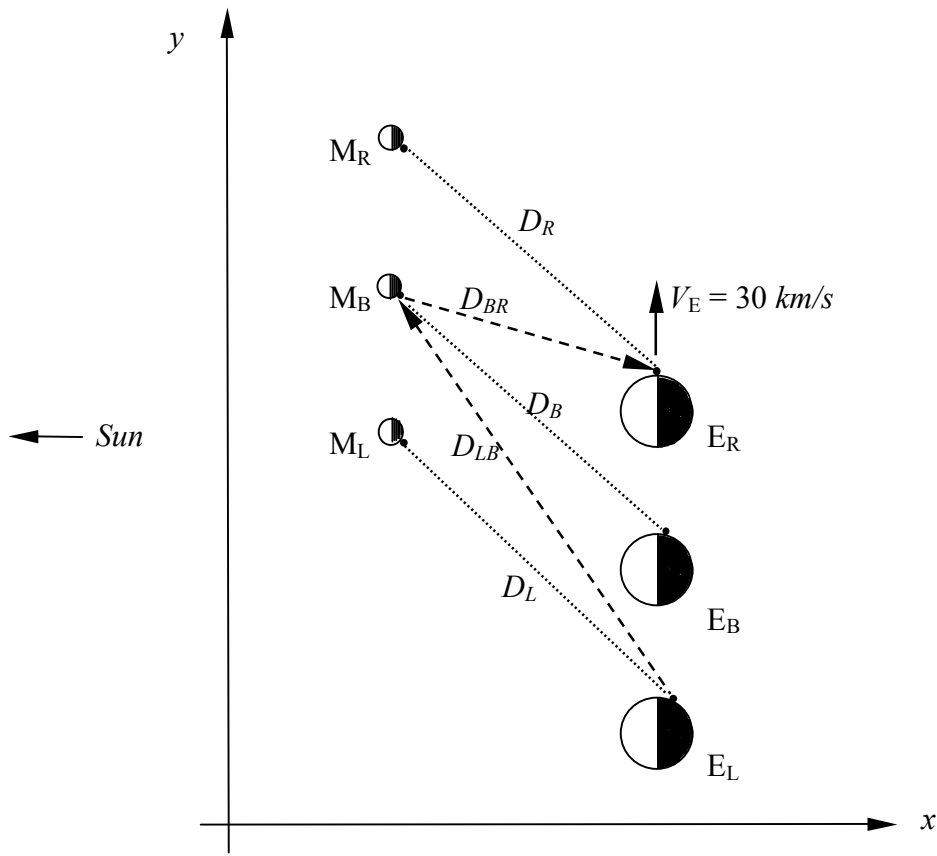

*Figure 2:* Schematic illustration of the *x,y* positions of the Earth (E) and Moon (M) in the non-rotating solar system barycentric J2000 inertial frame. The distances $D_L$, $D_B$ and $D_R$ are the actual separations of APO and A15RR calculated in the J2000 frame at the moments of launch (L), bounce (B) and receive (R). The distances $D_{LB}$ and $D_{BR}$ are the optical path lengths traveled from launch to bounce (LB) and from bounce to receive (BR), each derived from the position of APO and the position of A15RR at times separated by ~1.3 *sec*.

Table 1 gives observed times-of-flight (Murphy 2008) and corresponding modeled distances (Neumann 2008) for representative samples of the dataset at the beginning and end of the observing period, and near events $i$ = 1000 and 1100 (utilized in calculations that follow). The distance between APO and A15RR for each event $i$ = 1 - 2,636 during the observations is $D_{Xi}$, where $X$ denotes the event of launch ($L$), bounce ($B$) and receive ($R$), and the corresponding time-of-flight to the various distances is $T_{Xi}$, where $i$ is the individual shot number in the data set.

Table 1

Representative Measured Times and Modeled Distances

| Event $i$ | Launch Time (decimal days UTC) | Time-of-flight* $T_{LR}$ Predicted (sec) | Time-of-flight* $T_{LR}$ Measured (nsec) | Modeled Distances** $D_L$ | $D_{LB}$ | $D_B$ | $D_{BR}$ | $D_R$ | Modeled Times-of-Flight ** $T_{LB}$ | $T_{BR}$ | $T_{LB}+T_{BR}$ |
|---|---|---|---|---|---|---|---|---|---|---|---|
| 1 | 309.544444681877 | 2.637147914711 | 2637147908.590 | 395298.788295 | 395328.410401 | 395298.514285 | 395268.624438 | 395298.240353 | 1.318673635215 | 1.318474210708 | 2.637147845922 |
| 2 | 309.544448752171 | 2.637147427247 | 2637147394.494 | 395298.715217 | 395328.337302 | 395298.441217 | 395268.551399 | 395298.167293 | 1.318673391383 | 1.318473967076 | 2.637147358459 |
| 3 | 309.544449333939 | 2.637147357575 | 2637147392.839 | 395298.704772 | 395328.326854 | 395298.430773 | 395268.540959 | 395298.156851 | 1.318673356531 | 1.318473932254 | 2.637147288785 |
| 4 | 309.544451659074 | 2.637147079123 | 2637147055.369 | 395298.663028 | 395328.285098 | 395298.389034 | 395268.499238 | 395298.115118 | 1.318673217250 | 1.318473793086 | 2.637147010336 |
| 5 | 309.544460961609 | 2.637145965132 | 2637145958.456 | 395298.496023 | 395328.118046 | 395298.222051 | 395268.332323 | 395297.948157 | 1.318672660025 | 1.318473236318 | 2.637145896343 |
| 1000 | 309.548072538736 | 2.636720222301 | 2636720274.134 | 395234.670226 | 395264.274239 | 395234.404852 | 395204.541726 | 395234.139554 | 1.318459700008 | 1318260453789 | 2.636720153797 |
| 1001 | 309.548074866306 | 2.636719952288 | 2636719979.756 | 395234.629747 | 395264.233749 | 395234.364378 | 395204.501269 | 395234.099086 | 1.318459564946 | 1.318260318841 | 2.636719883787 |
| 1002 | 309.548074866306 | 2.636719952288 | 2636719952.842 | 395234.629747 | 395264.233749 | 395234.364378 | 395204.501269 | 395234.099086 | 1.318459564946 | 1.318260318841 | 2.636719883787 |
| 1003 | 309.548076029448 | 2.636719817359 | 2636719860.527 | 395234.609519 | 395264.213515 | 395234.344153 | 395204.481052 | 395234.078863 | 1.318459497454 | 1.318260251404 | 2.636719748858 |
| 1004 | 309.548077190871 | 2.636719682631 | 2636719682.872 | 395234.589320 | 395264.193311 | 395234.323957 | 395204.460865 | 395234.058671 | 1.318459430059 | 1.318260184068 | 2.636719614127 |
| 1100 | 309.548225444791 | 2.636702496314 | 2636702524.143 | 395232.012802 | 395261.616058 | 395231.747795 | 395201.885794 | 395231.482865 | 1.318450833271 | 1.318251594554 | 2.636702023620 |
| 1101 | 309.548228933934 | 2.636702092111 | 2636702092.242 | 395231.952205 | 395261.555444 | 395231.687207 | 395201.825230 | 395231.422285 | 1.318450631084 | 1.318251392536 | 2.636701821728 |
| 1102 | 309.548230676774 | 2.636701890217 | 2636701890.025 | 395231.921938 | 395261.525168 | 395231.656944 | 395201.794980 | 395231.392026 | 1.318450530095 | 1.318251291630 | 2.636701821728 |
| 1103 | 309.548231258447 | 2.636701822835 | 2636701822.699 | 395231.911836 | 395261.515064 | 395231.646843 | 395201.784884 | 395231.381927 | 1.318450496390 | 1.318251257956 | 2.636701754346 |
| 1104 | 309.548231840991 | 2.636701755352 | 2636701755.298 | 395231.901719 | 395261.504944 | 395231.636728 | 395201.774773 | 395231.371813 | 1.318450462632 | 1.318251224229 | 2.636701686862 |
| 2632 | 309.550275996740 | 2.636467151846 | 2636467151.863 | 395196.730624 | 395226.323770 | 395196.470571 | 395166.623631 | 395196.210595 | 1.318333110869 | 1.318133972640 | 2.636467083508 |
| 2633 | 309.550278324368 | 2.636466887217 | 2636466870.098 | 395196.690951 | 395226.284086 | 395196.430905 | 395166.583982 | 395196.170934 | 1.318332978496 | 1.318133840384 | 2.636466818880 |
| 2634 | 309.550278903734 | 2.636466821350 | 2636466848.788 | 395196.681077 | 395226.274208 | 395196.421031 | 395166.574113 | 395196.161062 | 1.318332945549 | 1.318133807464 | 2.636466753013 |
| 2635 | 309.550279487508 | 2.636466754982 | 2636466755.191 | 395196.671127 | 395226.264256 | 395196.411083 | 395166.564168 | 395196.151115 | 1.318332912350 | 1.318133774294 | 2.636466686644 |
| 2636 | 309.550280648937 | 2.636466622942 | 2636466622.769 | 395196.651332 | 395226.244455 | 395196.391291 | 395166.544385 | 395196.131326 | 1.318332846303 | 1.318133708302 | 2.636466554605 |

*Murphy (2008)  **Neumann (2008)

## 4. PRINCIPLE OF THE CALCULATION

Calculating the speed of light from modeled distance and measured time-of-flight would seem to be a simple exercise, yet there seems to be no agreement on what method would correctly apply, or how the results should be interpreted. Therefore, we must justify all details of this calculation, regardless how obvious or trivial they might be.

In general, we claim that the calculated speed of a light pulse in some reference frame is simply the distance the pulse travels in that frame from source to detector, divided by the time-of-flight of the pulse over that distance. The observatory (O) is moving along the line-of-sight at some speed $v_O$ in the local Earth-center/Moon-center stationary frame (S) due to the rotation of the Earth. The emitted light pulse reaches the detector after some elapsed time $T$ measured directly by the observer in frame O. The observer makes only one time-of-flight measurement, and this single time measurement may then be used for whatever speed calculation the observer might make, be it the speed of the pulse in frame O, in frame S, or in any other frame. But the optical path lengths are different in frame O and frame S when the observatory is moving (discussed in Section 2.3), and these lengths must be utilized appropriately.

### 4.1. Speed of Light in Frame S

First, consider the case where the observer is at rest in the local stationary frame (S). The distance between the observatory and the retro-reflector is $D$ at the moment the pulse is launched and the measurement starts, so the length of the optical path from the pulse to the detector at that moment is $2D$ (Figure 1). The stationary observer measures a time of flight $T$ and calculates the speed of the pulse $c_S$ in frame S to be

$$c_S = \frac{2D}{T} = c \tag{1}$$

The speed of light in frame S calculated by the stationary observer is $c$.

Now consider the case where the observer is moving in frame S at some slow speed $v_O$ toward the approaching pulse. The moving observer calculates the speed of the pulse in frame S. At the moment the measurement starts the optical path length from the pulse to the detector is $2D$. But motion of the observer in frame S during the measurement shortens the path traveled in frame S by $\Delta D$, where $\Delta D = v_O T$, so the pulse travels a shorter distance $2D - \Delta D$. The moving observer measures a correspondingly shorter time of flight $T - \Delta T$, where $\Delta T = \Delta D/c = v_O T/c$, so the speed of the pulse $c_S$ calculated in frame S is

$$c_S = \frac{2D - \Delta D}{T - \Delta T} = c \tag{2}$$

The pulse travels a shorter distance in frame S in a proportionally shorter time, so the speed of light in frame S calculated by the moving observer is also $c$.

### 4.2. Speed of Light in Frame O

Finally, the moving observer calculates the speed of the pulse in the observatory rest frame O. The pulse is a distance $2D$ from the observer in frame O at the start of the measurement. The observer is at rest in frame O, so the optical path length in frame O cannot change because the observer is moving in frame S. But the observer measures a shorter time-of-flight when moving toward the pulse in frame S than when at rest. The pulse is detected after an elapsed time $T - \Delta T$ on the moving observer's clock. The speed of the pulse in frame O, $c_O$, is then the full distance divided by the shorter time

$$c_O = \frac{2D}{T - \Delta T} = c + v_O \tag{3}$$

In this case the speed of light calculated in the observer's rest frame is found to exceed $c$ by $v_O$, the speed of the observatory along the line-of-sight.

*4.3. Predictable Objections*

Most physicists would agree that the moving observer does, indeed, detect the approaching pulse sooner when moving toward the pulse in the local stationary frame, but few would interpret that result as a violation of Lorentz invariance. There are several predictable objections to our approach, each of which can easily be refuted.

*4.3.1. The 'Shorter-Distance-in-a-Shorter-Time' Objection*

There is a standard argument that $c$ is invariant even though the moving observer detects the pulse sooner: The observer's motion toward the approaching pulse during the measurement shortens the optical path, so the pulse travels a shorter distance in a proportionally shorter time, and the calculated speed of light is $c$. But this familiar explanation is actually not correct; the 'shorter-distance-in-a-shorter-time' argument calculates the speed of the pulse in the local stationary frame (S), not in the frame of the moving observer (O). Motion of the observer during the measurement does not change the optical path length in frame O because the observer is at rest in frame O. In frame O the pulse and the observer are separated by $2D$ when the measurement starts, and they have to cover the full distance $2D$ between them to reach each other. But they do this in the shorter time $T - \Delta T$ measured when the observer is in motion, so the speed of light calculated in frame O is greater than $c$.

*4.3.2. 'Moving Retro-reflector' Objection*

It could still be argued that even though the observer is not moving in frame O during the observation, the retro-reflector is. This motion of the retro-reflector would act to shorten the optical path in frame O, the 'shorter-distance-in-a-shorter-time' argument would still apply in frame O, and the speed of light calculated in frame O would be $c$. The retro-reflector does indeed move in frame O when the observer is moving in frame S, but that motion does not change the total length of the optical path in frame O during the measurement (this is particularly hard to visualize, and completely counter-intuitive, but still true). The point in frame O at which the pulse originates is fixed in frame O. The retro-reflector folds the optical path (and frame O) back on itself. Motion of the retro-reflector toward the observer 'pushes' the folded segment of frame O continuously back, behind the observer. By the time the bounce occurs the distance from the origin of the pulse in frame O to the retro-reflector has increased from $D$ to $D+\Delta D$. But this motion simultaneously shortens the return leg by same the amount that it lengths the outbound leg, so that when the bounce occurs the length of the return leg has decreased to $D - \Delta D$ in frame O. (Note that after the bounce occurs the retro-reflector is behind the pulse for the rest of the experiment, so motion of the retro-reflector after the bounce can have no further affect on the length of the return leg). Thus, even with the retro-reflector moving in frame O during the measurement, the sum of the outbound and return legs in frame O is the full initial distance 2*D*.

*4.3.3. 'Classical Calculation' Objection*

Another likely objection might be that our result was obtained by a classical calculation, while problems involving propagating light should rightly be done using special relativity and the addition of velocities relation (ironically, the original 'shorter-distance-in-a-shorter-time' objection was based on a strictly classical calculation). The addition of velocities relation has only two variables, the speed of the observer in one frame and the speed of the thing being observed in another. When the speed of one of these things is *c* then the calculated relative speed of the two will always be *c*. The addition of velocities relation does not use any of the measured parameters that distinguish our experiment from all others (*i.e.*, the measured time of flight of the pulse, the rotation speed of the Earth, the actual distances between the laser source, the retro-reflector and the detector at various times during the measurement based on a database of lunar astrometry). In our experiment, the speed of the flash relative to the observer is unknown, and to be determined. In special relativity it is assumed *a priori* that the speed of light has the universal value *c* and is invariant for all observers. The purpose of this experiment is to test that assumption. Arbitrarily applying the relativistic addition of velocities relation would be applying the special theory of relativity to validate itself.

*4.3.4. 'Relativistic Effects are Ignored' Objection*

The objection could also be made that our calculation is wrong because relativistic time and distance differences between frame S and frame O are not accounted for. But the observatory is moving very slowly in the local stationary frame ($v_O \sim 200$ *m/sec* $\sim 10^{-6}c$), so the Lorentz factor is $\gamma \sim 10^{-12}$ in this calculation. Whatever relativistic length and time differences might exist would be second-order (millimeters in path length and picoseconds in time of flight), much smaller than our distance uncertainty and instrumental timing resolution (~1 meter and ~1 nanosecond) of the observation, or the first-order length and time changes that are actually observed (tens of meters and microseconds). Whatever relativistic effects might be invoked would make a negligible contribution to the uncertainty of the observation, and do not affect or disqualify a large, observed, first-order result.

*4.4. Principle of the Calculation - Summary*

The moving observer makes only a single time measurement, which can be used to calculate *c* in frame O, frame S, or in any other frame. With the observer in motion there is a large, easily measured difference between the optical path lengths in frame O and in frame S, which is orders of magnitude greater than any possible second-order affect of length contraction. Time dilation is also a second order effect, two orders of magnitude smaller at 200 *m/s* than our instrumental timing resolution, so any affect of time dilation would be undetectable, and completely negligible (as discussed in Sections 4.3.2). The moving observer calculates a first-order difference between the speeds of the pulse in frame O and in frame S, which is real and proportional to the observer's line-of-sight velocity.

Having established the advantages of the observational approach (Section 2) and the principles of the calculation, and addressed the obvious objections, we can proceed with the actual speed of light calculations in frame O and in frame S, using laser ranging data obtained with the observer in motion.

## 5. RESULTS

The speed of laser light pulses was calculated in the Earth-Moon stationary frame, and in the moving observatory frame, for each of the 2,636 events in the Murphy (2008) dataset. The speed of the observatory along the line-of-sight changes continuously during the observations, so the data were analyzed by making least squares fits to selected small groups of the slowly changing, measured times-of-flight.

### 5.1. Speed of Light in the Local Stationary Frame

The total distance traveled by a laser pulse in frame S is the sum of the length of the outgoing leg from launch to bounce $D_{LB}$ and the return leg from bounce to receive $D_{BR}$. The time-of-flight of the pulse is the measured interval $T_{LR}$ between the moments the shot is launched and received. Thus, according to eq. 2, the measured speed $c_S$ of the pulse in frame S is

$$c_S = \frac{D_{LB} + D_{BR}}{T_{LR}} \qquad (4)$$

Using the modeled geometrical distances provided by Neumann (2008) to compute the total optical path $D_{LB} + D_{BR}$ traveled by that pulse in frame S from launch to receive, and the time-of-flight $T_{LR}$ measured by Murphy (2008), and averaging speeds in frame S calculated using eq. 4 from a group of 100 events $i$ = 1000-1100, for instance, we obtain

$$\overline{c}_S = 299,792,449.7 \pm 3.2 \ m/s$$

This is less by 8.3 ± 3.2 *m/s* than the accepted universal value of *c* (299,792,458 *m/s*), for the average of those 100 events.

One might expect that a calculation using accurate time-of-flight measurements and properly calibrated distances would produce a measured speed-of-light result in frame S closer to the universal value of *c*. The ~8 *m/s* discrepancy could simply be due to imperfect light propagation corrections or other model-dependent factors. However, there is also the possibility that part of the difference between $c_S$ and $c$ is a real, measured, two-way velocity residual (discussed further in Secion 6). Rather than assume that we have a calibration error and adjust our distance model to give the expected result (which would be the usual approach) we accept this discrepancy as part of the measurement uncertainty to allow for the possibility that there is an underlying physical contribution to this result. Therefore, we conservatively estimate that our ability to measure the speed of light using this approach - without making any further assumptions or corrections - has an overall uncertainty of about ±10 *m/s*, which is still quite adequate for our purposes here. The speed of light calculated results in frame S for the complete data set is shown in Figure 3.

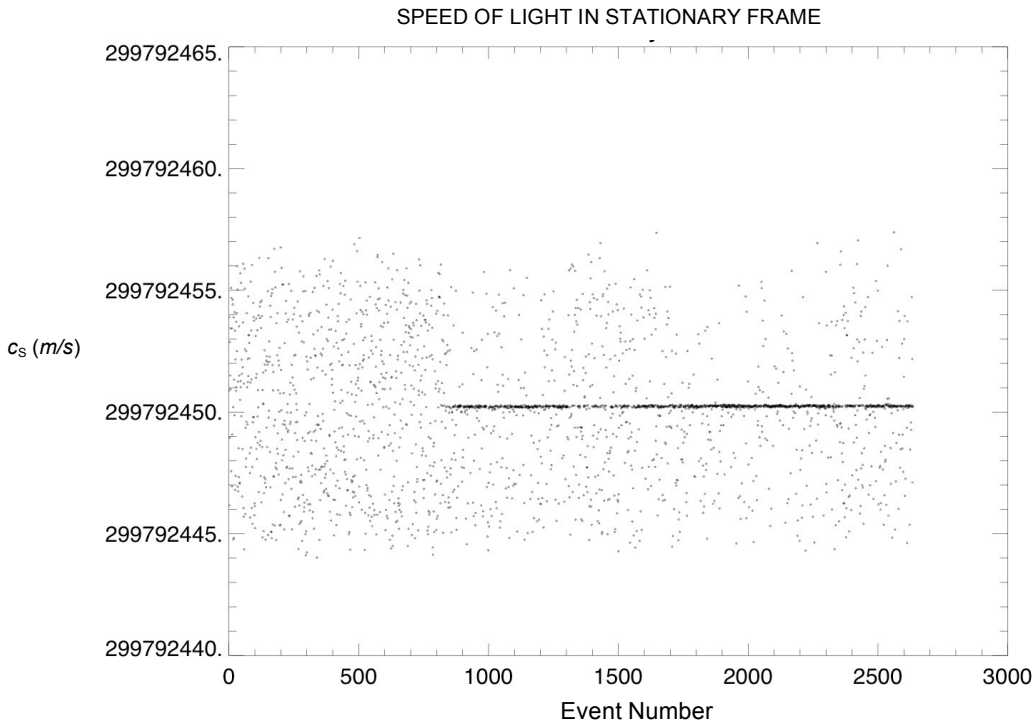

*Figure 3:* Speed of light calculated in the local Earth-Moon stationary frame (S).

*5.2. Speed of light in the observer's frame*

The calculated speed of light in frame O is just the total distance the pulse travels in frame O from the source to the detector divided by the time-of-flight measured by the observer. With the observer moving along the line-of-sight in frame S at $v_O$, the distance traveled by the pulse in frame O during the measurement is significantly different than the distance it travels in frame S, as discussed in Section 2.2. At the moment of launch the pulse is a distance $2D_L$ from the detector. In the case where the observer is at rest in frame O, motion of the observatory in frame S does not shorten the optical path in frame O during the measurement. (The apparent motion of the retro-reflector in frame O does not shorten the total optical path in frame O either, as discussed in Section 4.3.2). The pulse is received after the measured time $T_{LR}$ so, according to eq. 3, the speed of light $c_O$ in frame O is

$$c_O = \frac{2D_L}{T_{LR}}$$

In the case where the observer is in motion, the calculated value of $c_O$ changes because the speed of the observatory along the line-of-sight is changing continuously during the measurement (due to the changing zenith angle of the Moon). In the 3 *sec* time interval between events $i = 1000 - 1100$, for example, the Neumann (2008) model predicts that $v_O$ is decelerating at a rate of 0.8 *m/s*. By a least-squares fit to these 100 events, we obtain

$$\overline{c_O} = 299{,}972{,}655.8 \pm 3.3 \text{ } m/s$$

This result exceeds the universal value of *c* by $199.8 \pm 3.3$ *m/s* (Figure 3), which agrees to within our $\pm 10$ *m/s* experimental uncertainty with the predicted average speed of the APO of $v_O = 201$ *m/s* along the line-of-sight during that interval. Even with the overall measurement uncertainty estimated very conservatively (Section 5.2) this is a S/N = 20 result. Very similar results are obtained with other comparable groups of data. The calculated results in frame O for all of the shots in the dataset are shown in Figure 4.

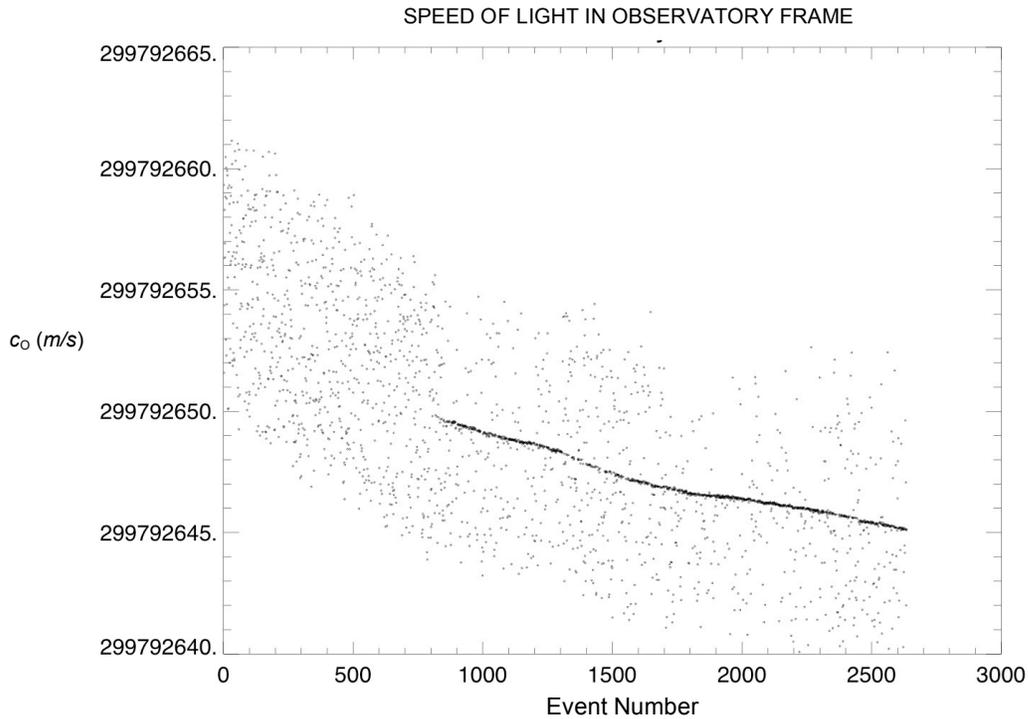

*Figure 4:* Speed of light calculated in the APO observatory frame (O), which was moving at 201 *m/s* toward the rising Moon along the line-of-sight in the Earth-Moon local stationary frame. Slight deviations in the calculated results from a smooth curve are due to the uneven time base of the plot caused by additional random noise events in the sample.

# 6. DISCUSSION

In familiar test theories of special relativity (*e.g.*, Mansouri and Sexl 1977) the observed speed of light is expressed as $c_O = c \pm k v_O$ where $v_O$ is the velocity of the observer along the line-of-sight in the local stationary frame, and the coefficient $k$ has the value $k = 1$ for classical relativity and $k = 0$ for special relativity. For the result obtained here we find $k = 0.95 \pm 0.05$. The observed speed of light measured by an observer moving at speed $v_O$ seems to follow the simple relation $c_O = c \pm v_O$. This result is not incompatible with nor does it preclude the idea that the speed of light itself is constant and invariant, or that light always propagates in free space at some unique speed, perhaps even at the nominal value $c = 299{,}792{,}458$ *m/s*. But it does imply that light travels at that unique speed only in one preferred or absolute reference frame. An observer moving relative to that preferred frame would measure the speed of light to be other than the nominal value.

Motion of an observer relative to a putative absolute frame would result in a difference in the time-of-flight between the outgoing and return legs of the optical path, but this would average out in a two-way measurement and appear only as a second order residual. For instance, the Earth's full 30 *km/s* orbital velocity along the line-of-sight would produce a second-order residual velocity of only ~3 *m/s*. So we cannot preclude the possibility that some part of the ~8 *m/s* difference between $c_O$ and $c$ measured here is a real second-order residual due to motion of the Earth-Moon system relative to an absolute frame. The present experiment cannot provide any insight into which physical system such a putative reference frame might be tied (the Earth, the Sun, the "fixed stars", the Galaxy, the cosmic background). Also unknown are what the perturbing affects might be, if any, of local gravitational and magnetic fields, solar wind plasma, cosmic rays, or dust particles in the zodiacal cloud.

If the results presented here are substantiated there would certainly be dramatic consequences for many areas of contemporary physics, astrophysics and cosmology. But the most obvious and troubling questions still remain: How could there be anything wrong with special relativity when it seems to work so well in the matter sector? And how could all of the successes of modern physics and cosmology have been possible if special relativity was not valid? The apparent successes of special relativity in the matter sector may not be fundamentally any more substantial than they are in the photon sector. The three predicted matter effects that seem to be so well supported experimentally all have alternate interpretations: Time dilation is taken as conclusively demonstrated by the extended half lives of energetic particles, but the causal relationship between time dilation and particle lifetimes is really only hypothetical and the underlying physical process is not understood. $E = mc^2$ can be derived without special relativity, as shown by Gould (2005). Mass increase was straightforwardly accounted for by the Ritz (1909) theory of electrodynamics and gravitation, where the difficulty of accelerating matter particles to speeds approaching $c$ was attributed not to mass increase, but rather to the finite propagation speed ($c$) of electromagnetic forces; a particle could not be accelerated to a speed faster than the force could act. It should also not be surprising that contemporary physics is fully compatible with special relativity because special relativity is at the very foundation of modern physics. Physics experiments today are analyzed with relativistic 'rods and clocks' and interpreted using the principles of special relativity. This kind of circular process results in internal consistency, but it can also lead to unrealistic dead ends. Consider the discomforting fact that the most interesting problems and concepts in physics and cosmology today involve phenomena that are virtual, unobserved, or unobservable. The obvious first places to look for difficulties in physics and cosmology in the photon sector resulting from the application of special relativity would be in the interpretation of the Hubble expansion and the problem of dark energy; in the matter sector it might be necessary to start by reconsidering the Dirac equation and its implications.

## 7. CONCLUSION

The most straightforward analysis and interpretation of two-way lunar laser ranging measurement of $c$ presented here suggests that light propagating between the Earth and the Moon obeys a classical rather than special relativistic addition of velocities law. On the face of it, this constitutes a first-order violation of local Lorentz invariance and implies that light propagates in an absolute reference frame, a conclusion that most physicists (except perhaps some contemporary field theorists) would be reluctant to accept. Rather than simply dismiss the present results and conclusions as implausible, which would be natural considering the strength of the prevailing view, it would be prudent to critically re-examine and improve the present experimental basis for special relativity in the photon sector. Ultimately, any concerns about the validity of a theory can only be resolved by experiment. We are now pursuing two new approaches to one-way measurements of the speed of light with slowly moving sources and detectors, both by one-way laser ranging outside the Earth's atmosphere (Gezari *et al.* 2010) and by direct optical pulse timing in the laboratory.

I am grateful to Robert A. Woodruff for many inspiring discussions and invaluable suggestions, to Don Goldsmith for astute advice and criticism, to Tom Murphy for providing lunar laser ranging data suitable for this study, and to Gregory Neumann for providing lunar distance modeling results. I deeply appreciate the objectivity and intellectual curiosity shown by Rick Lyon, Dale Fixsen, Randall Smith, Mark Mills, and Ben Oppenheimer, as well as the on-going collaborative contributions of Frank Varosi and Larry Woods. This effort was supported by Bill Danchi, Bill Oegerle and Jennifer Wisemann at Goddard, and by Mordecai-Mark Mac Low at the American Museum of Natural History. Special thanks to Lyn Stevens for her patience and encouragement. This research was supported by the National Aeronautics and Space Administration.